# INTERFACE EXCITONS IN Kr$_m$Ne$_N$ CLUSTERS : THE ROLE OF ELECTRON AFFINITY IN THE FORMATION OF ELECTRONIC STRUCTURE

A. Kanaev[1a], L. Museur[2], F. Edery[1], T. Laarmann[3], and T. Möller[3]

[1] *Laboratoire d'Ingénierie des Matériaux et des Hautes Pressions, C.N.R.S., Institut Galilée, Univerité Paris-Nord, 93430 Villetaneuse, France*

[2] *Laboratoire Physique des Lasers, C.N.R.S., Institut Galilée, Univerité Paris-Nord, 93430 Villetaneuse, France*

[3] *Hamburger Synchrotronstrahlungslabor HASYLAB at Deutsches Elektronen Synchrotron DESY, Hamburg, Notkestr. 85, 22603 Hamburg, Germany*

*Abstract*

The formation of the electronic structure of small $Kr_m$ clusters ($m \leq 150$) embedded inside $Ne_N$ clusters ($1200 \leq N \leq 7500$) has been investigated with the help of fluorescence excitation spectroscopy using synchrotron radiation. Electronically excited states, assigned to excitons at the Ne/Kr interface, $1i$ and $1'i$ were observed. The absorption bands, which are related to the lowest spin-orbit split atomic Kr $^3P_1$ and $^1P_1$ states, initially appear and shift towards lower energy when the krypton cluster size $m$ increases. The characteristic bulk $1t$ and $1't$ excitons appear in the spectra, when the cluster radius exceeds some critical value, $R_{Cl} > \delta_{1i}$. Kr clusters comprising up to 70 atoms do not exhibit bulk absorption bands. We suggest that this is due to the penetration of the interface excitons into the $Kr_m$ cluster volume, because of the negative electron affinity of surrounding Ne atoms. From the energy shift of the interface absorption bands with cluster size an unexpectedly large penetration depth of $\delta_{1i}$ =7.0±0.1 Å is estimated, which can be explained by the interplay between the electron affinities of the guest and the host cluster.

[a] corresponding author. E-mail : kanaev@limhp.univ-paris13.fr

**Introduction**

The formation of the electronic structure of solids is an important issue of cluster physics, which offers the possibility to study its evolution from atomic energy levels towards the band structure of bulk material as a function of cluster size. One of the interesting model systems are rare-gas clusters bound by weak pairwise Van-der-Waals forces. These clusters can be easily prepared in a supersonic expansion. Rare-gas clusters are transparent within the UV-visible spectral range and exhibit absorption bands in the vacuum ultraviolet (VUV). Rare-gas solids have fcc structure, while the corresponding clusters are icosahedrons with a total number of atoms $N = 1/3(10k^3 - 15k^2 + 11k - 3)$, where k is the number of the closed shells. The number of surface atoms is given by $N = 10k^3 - 20k^2 + 12$ (valid for k>1). One can see that even in clusters comprising 500 atoms almost 50% of the atoms belong to the surface. Moreover, due to the large number of surface atoms, clusters are ideal objects to study surface effects and, generally, interfaces in solids containing different materials.

Spectroscopy of free clusters of *He* [1], *Ne* [2], *Ar* [3, 6], *Kr* [4, 6] and *Xe* [5, 6] atoms has been studied since more than one decade. A special aspect of rare gas cluster is that their absorption bands split into electronically excited bulk and surface states. Since small clusters with less then two shells of atoms (k=2) have almost no bulk atoms, only surface excitons 1s and 1s' are observed in absorption. Bulk excitons appear, when the third shell of atoms is formed (k=3). Moreover, the experimental results show that tightly bound surface excitons have a very small penetration depth into the cluster, typically $\delta_{1s} \approx 0.8$ Å [7] and are therefore restricted to the surface atomic layer. Bulk excitons are delocalized within the rest of the cluster volume.

Recently, the so called "pick-up" technique allowed the growth of small guest rare-gas clusters inside large host rare-gas clusters and the investigation of embedded clusters with a shell-like geometric structure [8, 9]. This method has the advantage of controlling the temperature, the surface or bulk localization and the size of the guest cluster, as well as the size of the host cluster. Neon is a good solvent system for such experiments for several reasons: $Ne_{N>1000}$ clusters are soft enough and easily pick up atoms or molecules. They efficiently thermalize the dopant molecules at the characteristic cluster temperature of ~10K [10]. Moreover, the surounding neon cluster atoms do almost not perturb the energy levels of embedded molecules. Since neon clusters are transparent within the VUV spectral range, they are well-suited to study electronic properties of molecules and heavier rare-gas clusters made of $Ar$, $Kr$ and $Xe$ atoms. Recently, investigation of $Ne_N Ar_m$ clusters ($m \leq 100 << N \approx 7500$) gave insight into the tightly bound bulk excitons in small argon clusters [9]. This was possible because the dominant surface excitons of free argon clusters are suppressed inside neon. Additionally, these experiments allow the investigation of interface excitons as a function of the number of picked-up atoms. This is complimentary to earlier studies on surface excitons (bulk-vacuum interface), which were assigned by covering the surface of a rare gas solid with one-atomic layer of a different material.

In the present paper we report on the spectroscopy of small $Kr_m$ ($m \leq 150$) clusters embedded inside $Ne_N$ clusters ($1200 \leq N \leq 7500$). We observed new absorption bands assigned to the $Kr - Ne$ interface, as well as the known $Kr$ and $Ne$ bulk excitons. The energy shift of the observed bands as a function of the number of picked-up Kr atoms is discussed and from a detailed analysis, we gain information on the exciton creation, especially on peculiarities related to interface excitation.

**Experiment**

The measurements were performed at the experimental station CLULU at the synchrotron radiation laboratory HASYLAB in Hamburg [11]. Neon clusters were prepared in a supersonic expansion through a conical nozzle cooled down to 31 K. The average cluster size $N$ was determined using well-known scaling laws according to the formula [12, 13]:

$$N = 33 \cdot \left(\frac{\Gamma^*}{1000}\right)^{2,35}, \text{ with } \Gamma^* = K_{ch} \frac{p \cdot d_{eq}^{0.85}}{T^{2.2875}}, \quad K_{ch}(Ne) = 185,$$

and $p$ in mbar, $T$ in K and $d$ in μm are used. Depending on the geometry of the conical nozzle (200 μm, 4° or 100 μm, 15°), $Ne_N$ clusters with an average size of $1200 \leq N \leq 7500$ were prepared. The size distribution of the cluster beam has a width (FWHM) of approximately $\Delta N \approx N$.

Using a standard "pick-up" technique we have doped large cold neon clusters with Kr atoms from a cross-jet. The Poisson statistics determines the average number of picked-up atoms. The mean Kr cluster size $m$ has been estimated in the following way. The size of embedded $Ar_m$ clusters inside large neon clusters, which were prepared also in crossbeam experiments at CLULU, has been reported by Laarmann at al. [8]. The size determination based on theoretical and experimental work by Lewerenz at al. [14]. Moreover, Laarmann at al. [9] have shown that the absorption lineshape of tightly bound excitons in $Ar_m$ clusters changes with their size $m$ according to the Frenkel exciton model. By comparing VUV-fluorescence excitation spectra of $Ne_N Ar_m$ clusters in the range of 12.4 eV measured in the given experimental geometry with those from [9], one obtains a relation between the cross-jet pressure and the average number of embedded atoms. Since the probability for a Ne cluster to pick-up atoms is mainly depending on the Ne cluster size and the average cross-jet particle density along the beam axes, the calibration can also be used in the case of Kr doping. This calibration procedure has been applied recently in [15]. In the present work it resulted in $Kr_m$ cluster sizes $m \leq 150$.

Monochromatized synchrotron radiation ($\Delta\lambda$ = 0.25 nm bandpass) in the spectral range of 100-140 nm (*Al*-grating) or 40-100 nm (*Pt*-grating) was focused on the doped cluster beam 10 mm downstream from the nozzle. Fluorescence excitation spectra in the VUV-UV ($\lambda \leq 300$ nm) and in the UV-visible-IR ($200 \leq \lambda \leq 900$ nm) were recorded by two photomultipliers with CsI and GaAs(Cs) photocathodes, respectively. The background pressure was kept below $10^{-3}$ mbar during the experiments.

**Results and discussion**

Before presenting experimental results, some remarks on the cluster composition should be made. The pick-up of Kr atoms by large $Ne_N$ clusters results in a release of energy. Neon atoms are weakly bound and evaporate from the $Ne_N Kr_m$ cluster by heating. In fact, the binding energies per atom of rare-gas neon and krypton solids are 26.5 meV and 123.2 meV [16]. Therefore, doping decreases the initial cluster size by ~4.65 Ne-atoms per adsorbed Kr atom if the collision energy is neglected. Keeping in mind that the mean initial size of the $Ne_N$ cluster is *N*, the composition of the doped clusters can be written as $Ne_{N-4.65m}Kr_m$. With increasing number of picked-up Kr atoms the layer of Ne atoms, which cover the Kr surface becomes thinner. For example, with $N = 1200$ and $m \approx 10^2$ approximately 2 shells of neon atoms surround the krypton cluster.

It is well-known that after the absorption of VUV photons by rare gas clusters, energy relaxation processes are followed by VUV fluorescence of either atomic or molecular self-trapped excitons (aSTE/mSTE) or by fluorescence of desorbed electronically excited atoms, which emit in the VUV, as well as in the IR-visible spectral range [17]. Further, is well-known that tightly bound n=1 excitons in small rare-gas clusters decay radiatively in the VUV, whereas for excitons with n ≥ 2 the ejection of electronically excited atoms followed by infrared emission is observed. Doping of clusters change the situation, because of the energy transfer from excitons of the host-cluster to lower-lying energy levels of the embedded atoms or molecules. The appearance of n=1 excitons of the host cluster in the IR-visible fluorescence excitation spectrum is therefore a fingerprint of the pick-up process. In the case of Kr-doped Ne clusters excited at 17.64 eV the light is due to the transition (5p → 5s) of desorbed excited Kr atoms [8]. As an example, the IR-visible fluorescence excitation spectrum of $Ne_{990}Kr_{45}$ clusters is compared with that of $Ne_{1200}$ clusters in fig. 1. The n=1 excitons (17-18 eV) are not seen in pure neon clusters. They appear when Kr atoms are picked-up by the cluster. One important conclusion follows immediately from these spectra. Since the surface 1$s'$ exciton of neon does not appear in the IR fluorescence excitation spectrum, it follows that Kr clusters take interior sites of large neon clusters.

In the following we analyze the $n = 1,1'$ excitons of krypton in more detail, which firstly appear when Ne clusters are doped with a minimum number of Kr atoms. Since these excitons decay in the VUV spectral range, we measured VUV excitation spectra as a function of the mean krypton cluster size $m$. The results of $Ne_{7500-4.65m}Kr_m$ clusters are shown in fig. 2. Because of the presence of free atoms in the interaction volume, two narrow lines at 10.033 eV ($^3P_1$) and 10.644 eV ($^1P_1$) are always observed in the spectra. Except for these atomic lines, the spectra are completely different from that earlier reported for free krypton clusters [4]. Absorption bands identical to those in [4] are only observed if less that three shells of neon atoms remain on the surface of the embedded Kr cluster (spectra of $Kr_{150}$ and $Ne_{1200-4.65m}Kr_{m=100}$ clusters are also shown in the uppermost frame of fig. 2).

The main features in the fluorescence excitation spectra of embedded $Kr_m$ clusters ($m < 150$) are two new broad bands, which exhibit a red shift with increasing cluster size $m$. Because of the characteristic energy gap between these bands and their appearance in very small krypton clusters, we assign these bands to the interface excitons $1i$ and $1'i$ related to spin-orbit split $^3P_1$ and $^1P_1$ krypton atomic states. More arguments to support this assignment will be given in the following. Earlier studies on rare-gas alloys discovered excitonic bands, which were assigned in the low concentration limit to impurity atoms (see in [18]). Krypton atoms in neon solids exhibit an absorption band at ~10.68 eV [19, 20]. Recent studies on light rare-gas (*He*, *Ne*, *Ar*) clusters doped with heavy rare-gas atoms *Kr* and *Xe* confirmed this assignment [21]. In the case of very small $Ne_N Kr$ clusters ($N=12$) one band at 10.78 eV has been firmly assigned to perturbed electronically excited krypton atoms ($^3P_1$) surrounded by Ne cluster atoms. For large neon clusters ($N=10^3$), which is close to the size of our host clusters, this band shifts to 10.73 eV. The low-energy band ($1i$) observed in the present work, indeed firstly appears in this energy range (see fig. 2). In the limit of large Kr clusters both $1i$ and $1'i$ bands converge towards the position of the respective $1l$ and $1'l$ excitons.

Two absorption bands appear at 10.102 eV and 10.793 eV for $Kr_m$ clusters as large as $m \geq 70$. Due to their spectral positions, they are assigned to $1t$ and $1't$ bulk excitons. This cluster size corresponds to clusters composed of more than three complete shells of atoms, which is considerably larger than what is required for the formation of bulk excitons in free krypton clusters. One extra band appeared at 11.155 eV in large krypton clusters ($m \approx 80$), which can be assigned to the bulk $2t$ exciton. The longitudinal $1l$ and $1'l$ excitons are usually less intense than transverse excitons in free clusters. They are superimposed by the broad interfaces excitonic bands and therefore not clearly visible in the excitation spectra. Another line can be seen at 10.689 eV ($m \geq 70$), which assignment is not straightforward.

This energy correspond to the 1's exciton in free krypton clusters but also to the excitation of single $Kr(^3P_1)$ atom in the neon matrix [20]. Nevertheless, in the last case the intensity should decreases when the krypton pressure increases which is not our case. We have observed that the intensity ratio between the lines at 10.689 eV and 10.644 eV ($^1P_1$ of free Kr atoms) is independent on the host neon cluster size $N$. Because of that, we assign the line at 10.689 eV to the surface exciton 1's of free very small krypton clusters formed out of the neon clusters due to collisions. For the same reason, the shoulder at 9.94 eV is assigned to the surface exciton 1s of very small free krypton clusters.

Now, we will discuss the nature of the energy shift of the 1i and 1'i interface excitons. Since the radius of the first exciton in krypton $r_{n=1}(Kr)$=2.5 Å is smaller than the nearest neighbor distance $d_{Kr-Kr}$=3.98 Å [16], we can understand the energy shift towards lower energy with increasing cluster with the help of the Frenkel exciton model taking the resonant excitation transfer into account. Recently, this model has been successfully applied to explain the red shift of n=1 excitonic bands in $Ne_N Ar_m$ clusters ($m << N \approx 7500$) [9]. In particular, it was shown that the shift of the interface excitation is proportional to the logarithm of the number of *surface* atoms $m_S$ of the embedded argon cluster. As our results show (see fig. 3), in small $Ne_N Kr_m$ clusters ($m < 80$) the energy shift of the interface exciton bands 1i and 1'i is proportional to the logarithm of the *total number* of krypton atoms: $\Delta E \propto \ln(m)$, which indicates that in this range of cluster sizes all atoms participate in the resonant energy transfer [9, 22]. This experimental result suggests a large penetration depth of the interface excitons inside the krypton cluster, which is in contrast to $Ne_N Ar_m$ clusters, where the interface exciton is localized within the surface *Ar*-atomic layer.

To estimate the exciton penetration depth we proceed as following. Since excited atoms have no permanent dipole moment, the energy shift of the exciton band is mainly described by the resonance interaction term [9, 22]

$$L_f(\mathbf{k}) = \sum_p M_{np}^f \exp(i\mathbf{k}\cdot(\mathbf{n}-\mathbf{p})) \quad (1)$$

where $M_{np}^f$ is the matrix element of the excitation transfer between atoms in positions with radius vectors $\mathbf{n}$ and $\mathbf{p}$. In the case of dipole-dipole interaction $M_{np}^f$ is expressed as

$$M_{np}^f = \frac{1}{r_{np}^5}\left[(\mathbf{d}_n\mathbf{d}_p)r_{np}^2 - 3(\mathbf{d}_n\mathbf{r}_{np})(\mathbf{d}_p\mathbf{r}_{np})\right] \quad (2)$$

where $d_n = d_p = d/\sqrt{\varepsilon}$ and $\varepsilon$ is the dielectric constant of the solid krypton. Since the radius of the krypton cluster $R_{cl} = m^{1/3} r_0$ is small compared to the wavelength of the excitation light, the term $L_f(\mathbf{k})$ can be calculated by replacing the summation in (1) by an integration of the different contributions of $M_{np}^f$ [22]. We obtain

$$L_f(m) \propto C(R_{cl},\delta) \int_{Cluster} \frac{\rho_i(r)}{r^3} dV \quad (3)$$

where $\rho_i(r) = \exp\left(-\frac{(R_{cl}-r)}{\delta}\right)$ is the density probability function of a surface exciton [23], $\delta$ the penetration depth of the surface exciton inside the cluster and $C(R_{cl}, \delta)$ is a normalization constant $C(R_{cl},\delta)\int_{cluster}\rho_i(r)\,dV = 1$. We have used the expression (3) to fit the experimentally observed energy shift. The best fit for both $1i_S$ and $1'i_S$ interface excitons results in $\delta_{1i} \approx 7.0\pm0.1$ Å. These curves are shown by solid lines in fig. 3. We have to remark that the bulk $n = 1,1'$ excitons appear only, if the krypton cluster radius exceeds the exciton penetration depth ($R_{cl} > \delta_{1i}$), which corresponds to an average number of picked-up Kr atoms $m > 70$.

Surface excitons have larger binding energies than bulk excitons. Qualitatively, this is explained by the fact that the dielectric screening at the surface is smaller than in the bulk material. As a result, the surface exciton absorption band is red shifted with respect to the bulk one in optical spectra. Contrary to the case of the solid-vacuum boundary, in the case of more complex interfaces, like $Kr-Ne$, one has to consider a perturbation of the excited electronic orbital between both solid phases forming the interface, which results in the energy shift of the band. Here, the electron affinity of the solids plays a key role. It is known, that the sign and the value of the electron affinity $V_0$ are determined by the interplay of polarization and short-range repulsion of an excess electron. In earlier studies, the sign of $V_0$ has been evoked to explain desorption processes of electronically excited atoms from the surface of rare-gas solids [24]. In the so-called "cavity-ejection mechanism", the excited atom polarizes the solid and its remote excited electron cloud undergoes short-range repulsion.

The electron affinity of the respective material may also be useful for an explanation of the properties of the interface exciton formation. In the case of the $Kr-Ne$ interface, the electron affinity of bulk neon is negative $V_e=-1.3$ eV, whereas that of bulk krypton is positive $V_e=+0.3$ eV [24]. This is also valid for krypton clusters ($m \geq 16$), where the electron affinity change its sign and becomes positive [25]. Because of the stronger repulsion from the neon phase, the interface exciton is pushed into the krypton condensed phase. This may explain the experimental finding of the present work: an extremely large value of $\delta_{Ne-Kr} \equiv \delta_{1i}$ =7.0 Å. The strong perturbation of the exciton at the interface is also evident from its large width, which is generally much narrower for surface excitons compared to bulk excitons [16]. When the size of the embedded krypton cluster increases, the interface exciton width becomes narrower.

To shed light on the correlation between the electron affinity of embedded clusters and the interface exciton penetration depth, we investigated $Ne_N Xe_m$ clusters in another set of measurements. We like to note, that the electron affinity of bulk xenon is $V_e = +0.4$ eV. We found that the exciton formation of $Ne_N Xe_m$ clusters is very similar to that of $Ne_N Kr_m$ clusters and an exciton penetration depth $\delta_{Ne-Xe} \equiv \delta_{1i} = 6.5$ Å was derived from the experimental data [26]. On the other hand, in earlier studied $Ne_N Ar_m$ clusters no bulk delocalization of the interface exciton has been observed [9], and a small penetration depth $\delta_{Ar-Ne} \approx 0.54 \pm 0.06$ Å has been reported in ref. [27]. The electron affinity of bulk argon is negative, $V_e = -0.4$ eV [24], and this value is expected to be even higher in small clusters, where polarization forces are weaker [25]. Because of that, the interface exciton is expected to be confined within the uppermost $Ar$-atomic layer in agreement with the experiments.

**Conclusion**

In conclusion, we have experimentally studied the formation of tightly bound $n = 1,1'$ excitons in small $Kr_m$ clusters ($m \leq 150$) embedded inside large $Ne_N$ clusters ($1200 \leq N \leq 7500$) with fluorescence excitation spectroscopy. We have observed absorption bands due to excitons at the $Kr-Ne$ interface ($1i$ and $1'i$). Bulk excitons of Kr$_m$ clusters ($n = 1,1'$) only appear in sufficiently large clusters with $m > 70$. The interface excitations shift towards lower energy with increasing Kr cluster size. This can be explained with the help of the Frenkel-exciton model taking the resonant excitation transfer into account. We have determined the penetration depth of the interface exciton into the bulk material: $\delta_{Kr-Ne} \approx 7.0 \pm 0.1$ Å. This value is unexpectedly large and may be explained by the interplay between electron affinities of the adjacent condensed solid phases composing the interface.

This work was supported by the IHP-Contract HPRI-CT-1999-00040 of the European Commission.

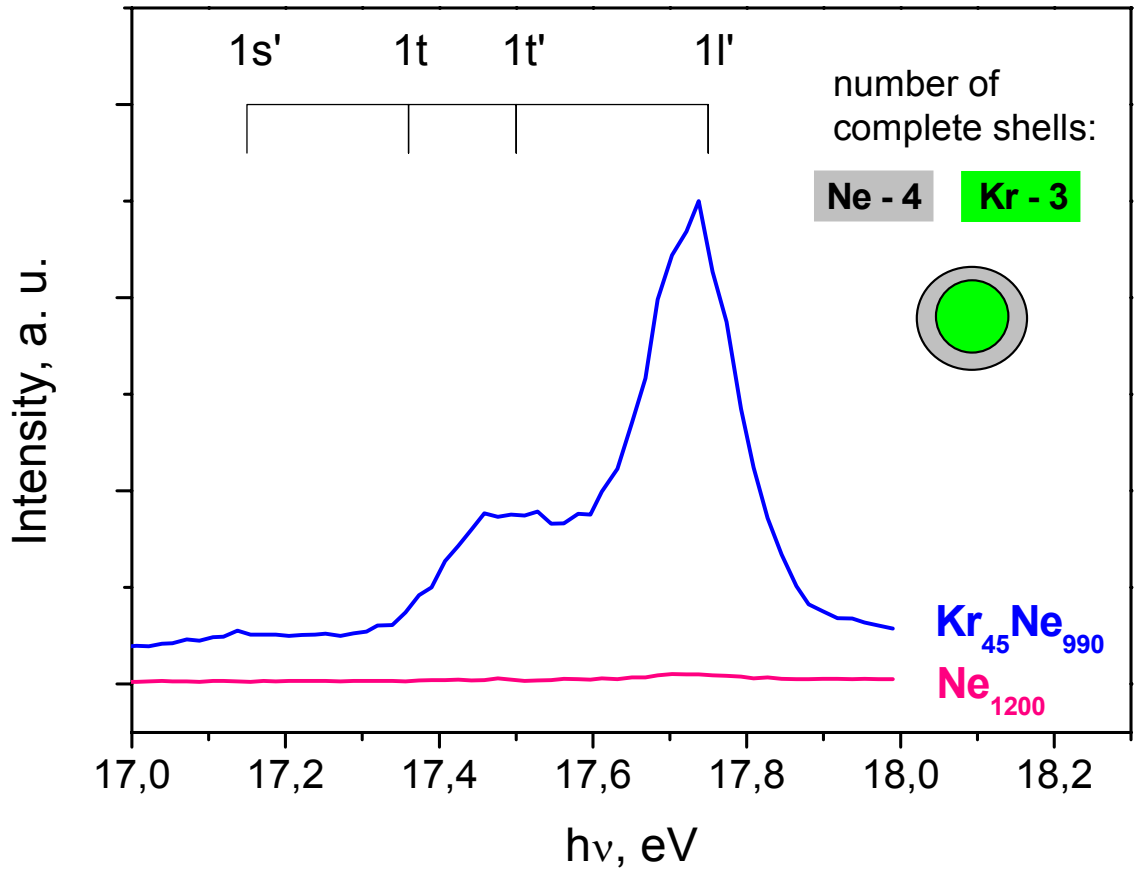

Fig.1  IR-visible fluorescence excitation spectra of $Ne_N Kr_m$ clusters.

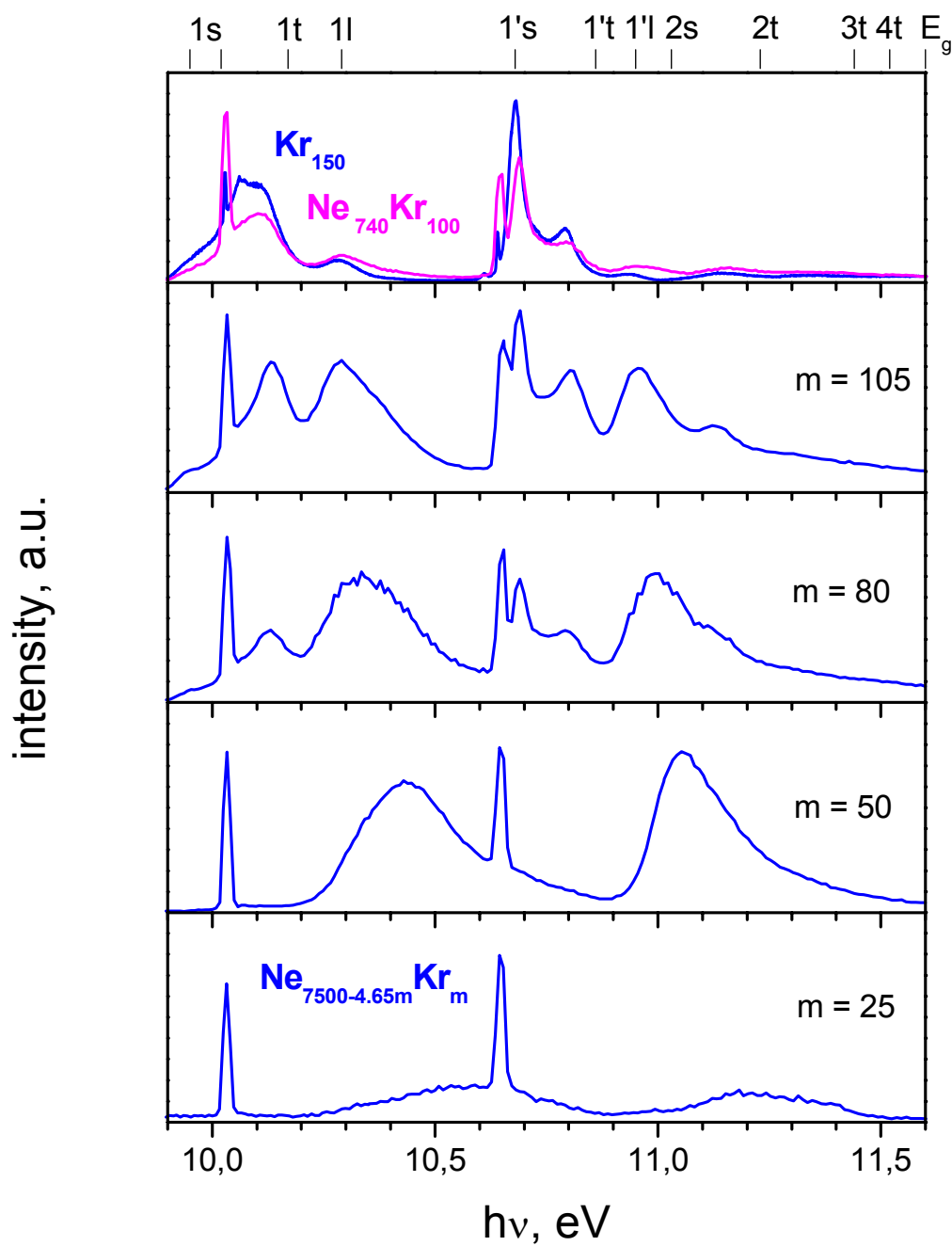

Fig.2  VUV fluorescence excitation spectra of $Kr_{150}$ and $Ne_N Kr_m$ ($N = 1200$ and 7500) clusters in the energy range of the krypton cluster absorption.

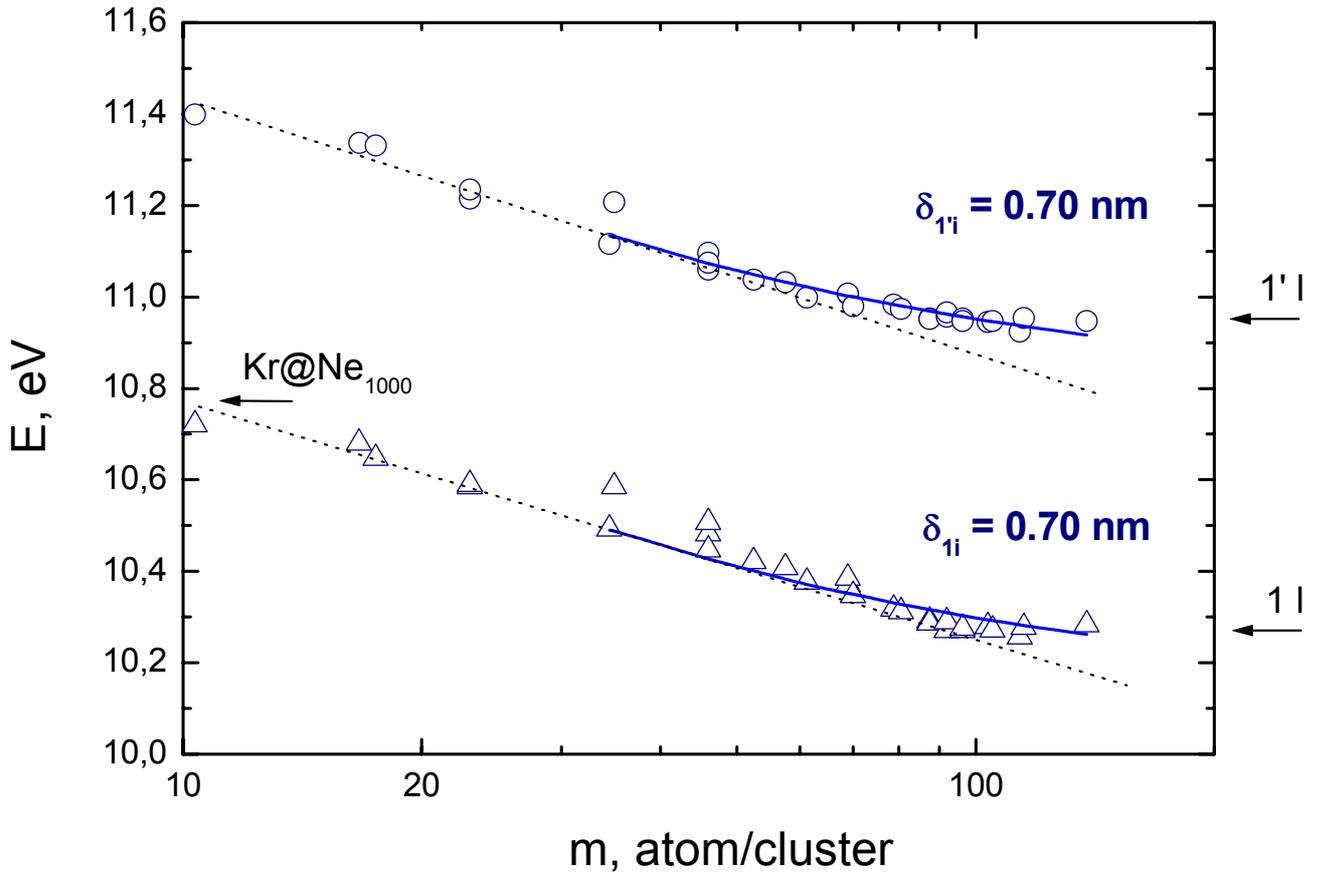

Fig.3  Energy shift of the 1*i* and 1'*i* interface (*Kr – Ne*) excitons in $Ne_N Kr_m$ clusters versus the cluster size *m*. The full lines represent the fit obtained with the expression (3). The energetic positions of 1*l* and 1'*l* excitons as well as the energy position of perturbed electronically excited *Kr* atoms ($^3P_1$) inside bulk neon are indicated. A pure logarithmic dependence of $\Delta \nu \propto \ln(m)$ is valid for small *m*; it is given by the dotted lines.